\documentclass[10pt, conference, letterpaper]{IEEEtran}

\usepackage{times}  
\usepackage[bookmarks=true]{hyperref}
\usepackage{subcaption}
\usepackage[noadjust]{cite}

\usepackage{color, colortbl}
\definecolor{Gray}{gray}{0.9}
\definecolor{Vio}{rgb}{0.5,0,1}
\definecolor{Org}{rgb}{1,0.2,1}
\definecolor{LightCyan}{rgb}{0.88,1,1}
\definecolor{LightRed}{rgb}{1,0.88,0.88}
\definecolor{LightBlue}{rgb}{0.12,0.7,0.8}
\definecolor{LighterBlue}{rgb}{0.12,0.9,0.9}
\definecolor{Red}{rgb}{1,0,0}
\definecolor{Green}{rgb}{0,0.5,0}
\definecolor{LightGreen}{rgb}{0,1,0}
\usepackage[table]{xcolor}% http://ctan.org/pkg/xcolor
\usepackage{amssymb}% http://ctan.org/pkg/amssymb
\usepackage{pifont}% http://ctan.org/pkg/pifont
\newcommand{\cmark}{\ding{51}}%
\newcommand{\xmark}{\ding{55}}%

\newtheorem{thm}{\underline{Theorem}}
\newtheorem{coro}{\underline{Corollary}}
\newtheorem{define}{\underline{Definition}}
\newtheorem{lem}{\underline{Lemma}}

\usepackage{amsmath} 
\usepackage{amsfonts}
\usepackage{mathtools}% http://ctan.org/pkg/mathtools
\DeclarePairedDelimiter\ceil{\lceil}{\rceil}
\DeclarePairedDelimiter\floor{\lfloor}{\rfloor}

\usepackage{tabulary}

\usepackage{algorithmic}
\usepackage[linesnumbered,ruled,vlined]{algorithm2e}
\SetKwProg{Fn}{Function}{}{}

\SetCommentSty{mycommfont}
\makeatletter
\newcommand{\removelatexerror}{\let\@latex@error\@gobble}
\makeatother

\usepackage{dblfloatfix}    % To enable figures at the bottom of page

\begin{document}

\title{SharpEdge: An Asynchronous and Core-Agnostic Solution to Guarantee Bounded-Delays}
%\title{Guaranteed Bounded Delays with a Simple Asynchronous \& Core-Agnostic Solution}

\author{\IEEEauthorblockN{Anonymous Author(s)}
\IEEEauthorblockA{%\textit{Tandon School of Engineering} \\
\textit{ }\\
 }

\author{\IEEEauthorblockN{Soheil Abbasloo}
\IEEEauthorblockA{%\textit{Tandon School of Engineering} \\
\textit{New York University}\\
ab.soheil@nyu.edu}
\and
%\IEEEauthorblockN{2\textsuperscript{nd} Yang Xu}
%\IEEEauthorblockA{%\textit{dept. name of organization (of Aff.)} \\
%\textit{Fudan University}\\
%email address}

%\and
\IEEEauthorblockN{Jonathan H. Chao}
\IEEEauthorblockA{%\textit{Tandon School of Engineering} \\
\textit{New York University}\\
chao@nyu.edu}
}
%\and
%\IEEEauthorblockN{4\textsuperscript{th} xxx}
%\IEEEauthorblockA{%\textit{Tandon School of Engineering} \\
%\textit{Huawei}\\
%chao@nyu.edu}
%\and
%\IEEEauthorblockN{5\textsuperscript{th} xxx}
%\IEEEauthorblockA{%\textit{Tandon School of Engineering} \\
%\textit{Huawei}\\
%chao@nyu.edu}
%\and
%\IEEEauthorblockN{6\textsuperscript{th} xxx}
%\IEEEauthorblockA{%\textit{Tandon School of Engineering} \\
%\textit{Huawei}\\
%chao@nyu.edu}
%\and
%\IEEEauthorblockN{7\textsuperscript{th} xxx}
%\IEEEauthorblockA{%\textit{Tandon School of Engineering} \\
%\textit{Huawei}\\
%chao@nyu.edu}
}
\maketitle

\begin{abstract}
What are the key properties that a network should have to provide bounded-delay guarantees for the packets? In this paper, we attempt to answer this question. To that end, we explore the theory of bounded-delay networks and provide the necessary and the sufficient conditions required to have deterministic bounded-delays in the network. 

We prove that as long as a network is work-conserving, independent of the packet scheduling and queue management algorithms used in the switches, it is sufficient to shape the traffic~\textit{properly} at the edge of the network to meet hard bounded-delays in the network. 

Using the derived theorems, we present SharpEdge, a novel design to meet deterministic bounded-delays in the network. To the best of our knowledge, SharpEdge is the first scheme that can meet all following key properties: 1) it supports coexistence of different classes of traffic, while it can guarantee their different required bounded-delays 2) it does not require any changes in the core of the network, 3) it supports both periodic and bursty traffic patterns, and 4) it does not require any time synchronization between network devices.
\end{abstract}


\section{Introduction}
\label{intro}
\subsection{Setting the Context}
The goal of satisfying real-time communications in a network and providing delay guarantees is not new for network protocol/system designers. The early 90s saw the first wave of queue management and scheduling proposals for supporting real-time traffic and providing performance guarantees in the network. 
%First wave of delay-centric network protocol/system designs happened during the early 90s. 
Schemes such as GPS and its packetized version PGPS~\cite{gps1}~\cite{gps_multi}, Stop-and-Go~\cite{sg1}~\cite{sg2}, HRR~\cite{hrr}, D-EDD~\cite{edd_d}, RCSP~\cite{rcsp} are some of the popular schemes among the vast number of proposals that consider the delay as a key performance metric for designing queue management and scheduling techniques. 

The main target of the schemes in the early 90s was the Internet. However, the Internet has grown based on the simple commercial switches and routers, which usually for cost concern do not equip with complex queue management and scheduling algorithms. Hence the best-effort traffic became the king of the traffic on such an environment and the need for new approaches with delay guarantees dimmed. 

However, another wave of new proposals with the theme of performance guarantees in the network started in early 2010s with the main focus on the datacenter networks. The key motivation was the new delay requirements of emerging online data-intensive (OLDI) applications such as Web search, retail, advertisement, social networking, and recommendation systems in datacenters at that time. Another important factor that encouraged new designs was the single authority nature of datacenter networks. In contrast with the Internet where the entire path of the traffic is not in the control of a single entity, datacenters are owned and managed by a single entity. This property gave researchers more freedom in exploring design space~\cite{d3,pfabric,qjump,pdq,hyline,no_sch}.
%considering delay as the key performance metrics in datacenter %For instance, previously, the designs that required changes in all network devices on the path of traffic to guarantee the delay performance were abandoned in the Internet because all different entities involved in the path of the traffic were required to implement the design which faced logistic/marketing issues, while this was not an issue in a single authority network such as datacenter. 

Recently, the ultra-low latency requirements of emerging applications such as virtual reality, augmented reality, vehicle to vehicle communications, real-time online gaming, real-time industrial control, automated driving, tactile Internet, etc. have created another wave of delay-oriented network designs and motivated the community toward rethinking of the currently dominant throughput-oriented design structures which is used as the base for most of the existing network protocols, architectures, and devices (e.g.~\cite{sprout,c2tcp,deepcc,natcp}). The emerging 5G technology which holds the promise of improved latency in the network is another motivating factor for the recent wave of delay-centric designs to provide delay guarantees for the packets in the network. 

\subsection{A Brief Background on the Efforts Done by IEEE to Propose Solutions with Delay-Performance Guarantees}
%\subsection{IEEE and Time-Sensitive Networks}
Among various efforts and studies done in the network community, IEEE started to work on delay-sensitive networks in 2007 under the task group named IEEE AVB (audio video bridging). IEEE AVB focused on providing bounded-delay communication for audio/video devices in a small network. However, it did not get success in the market. So, in 2012, the ABV task group was changed to time-sensitive network (TSN) task group to broaden the ABV's objective and cover more time-sensitive applications. IEEE TSN has focused on layer 2 solutions to provide guaranteed bounded delay communications. IEEE TSN has released different standards addressing bounded-delay communications in the network including IEEE 802.1Qav~\cite{cbs} (released in 2009), IEEE 802.1Qbv~\cite{tas}    (released in 2015), and IEEE 802.1Qch~\cite{cyclic} (released in 2017). Currently, the IEEE TSN task group is working on a newer version under the name of IEEE 802.1Qcr~\cite{ats}. However, all these efforts and standards (including the latest ongoing work) have issues which prevent them from becoming the ultimate solutions. %(detailed in section~\ref{sec_related_ieee}).

\subsubsection{IEEE 802.1Qav~\cite{cbs}} 
802.1Qav is the first proposal by IEEE AVB targeting audio/video devices. In 802.1Qav, flows with reserved rates are put into two classes/queues and traffic at the ingress of each queue is shaped using a credit based shaper (CBS). The key idea for CBS is to spread the packets of each class through time to avoid the burstiness of traffic in the network while making room for sending traffic of the other class. Although CBS is a simple approach, it is shown that under high link load, CBS cannot provide delay guarantees~\cite{cbs_issue}.  Moreover, CBS is a non-work-conserving approach which can lead to underutilized links.
\subsubsection{IEEE 802.1Qbv~\cite{tas}} 
To overcome the problems of CBS, TSN proposed a new shaper in 802.1Qbv standard called Time-Aware Shaper (TAS). TAS uses a TDMA flavor approach. In particular, it is considered that all devices are synchronized using a global clock. Then, the delay-sensitive traffic will be sent throughout the network in time-triggered guarded windows. In other words, when delay-sensitive traffic is to be sent, all other queues will be disabled for a particular time window throughout the network. TAS is able to meet delay-bounds in the network. However, it requires a fully synchronized network and a complete change of all switches in the network. 
\subsubsection{IEEE 802.1Qch~\cite{cyclic}} 
IEEE 802.1Qch or Cyclic Queuing and Forwarding (CQF) is the recent TDMA flavor approach proposed by TSN. This approach is very similar to Stop-and-Go~\cite{sg1,sg2}. In CQF time axis is divided into a sequence of Cycles (similar to Frames idea in Stop-and-Go). A higher priority traffic received in an input cycle will be served at the next cycle at the output. CQF can bound traffic's delay to $2\times T$ on each hop, where $T$ is the duration of a Cycle. Therefore, overall bounded-delay of network will depend on the number of hops which is a disadvantage.

\subsubsection{IEEE 802.1Qcr~\cite{ats}} 
\label{sec_802.1Qcr}
Both 802.1Qbv and 802.1Qch require a fully synchronized network. In addition, due to the use of enforced periodic cycles to transmit packets, they will cause underutilized links.
So, to overcome these issues, TSN is currently working on an asynchronous solution under the name of 802.1Qcr amendment. Basically, the main approach~\cite{ubs} which is proposed to be used as the base in 802.1Qcr is to employ the RCSP~\cite{rcsp} framework and choose leaky buckets~\cite{leaky} as the rate regulators in RCSP. However, even with this latest TSN effort, multiple issues still remain. First, 802.1Qcr requires per-flow shaping (per-flow management) at each hop in the network which means all switches in the network are required to be changed. Second, 802.1Qcr simply follows the greedy delay bound approach in which the end-to-end delay is calculated by considering a maximum delay per hop and then multiplying that by the number of hops. So, similar to IEEE 802.1Qch, the overall delay will depend on number of hops. Moreover, RCSP (therefore, 802.1Qcr) is a non-work-conserving design which leads to underutilized links.

\subsection{Motivations and Contributions}
Motivated by the need for a solution addressing ultra-low latency requirements of emerging applications in the network and the lack of a simple deployment-friendly solution which can guarantee deterministic delays in the network, in this paper, we investigate the theory behind bounded-delay networks.

In contrast to most of the studies that start from proposing a solution and later prove that the proposed solution works, we start from the theory behind the bounded-delay networks and without tying it to any specific design/scheme, we derive required theorems to answer these key questions: 
\begin{itemize}
\item Necessary condition: What are the minimal common properties of the networks that have guaranteed bounded-delays?
\item Sufficient condition: Under which conditions will a network have a guaranteed bounded-delay? 
\end{itemize}

Having these questions answered, we use the derived theorem to propose a simple and deployment-friendly solution called SharpEdge to achieve deterministic bounded-delay guarantees in the network. It is worth mentioning that SharpEdge is an example of the solutions that can satisfy the sufficient conditions that we derive and it is not the only one. The provided theorems can be employed as the base to propose other schemes. Also, they can be used to examine whether a proposed scheme is able to bound the delay in a network.

\textbf{The Key Result}: We prove that independent of the algorithms used for serving packets in the network, it is sufficient to shape the traffic \textit{``properly''} only at the edge of the network to guarantee bounded-delay for the packets in the network. The only requirement of the network is that it needs to employ \textit{``work-conserving''}\footnote{A work-conserving network is a network that keeps serving packets when there are packets in the network. In contrast, a non-work conserving network is a network which, in some cases, may leave the output links idle despite the presence of packets at the input.} (queue management/scheduling) algorithms similar to the ones that are already used in the commercial switches to serve incoming packets. 

\textbf{Properties of SharpEdge}:
SharpEdge has 4 main properties which distinguish it from the designs and proposals existed in the literature:
\begin{enumerate}
\item It can guarantee various bounded-delays in a network.
\item It only introduces a novel Shaper employed at the edge of the network (either at the end-hosts or at the edge switches). Therefore, it 
%is a deployment-friendly approach which 
does not require any changes in the core network.
\item It supports traffics with burstiness.
\item It does not require any time synchronization among network devices.
\end{enumerate}

These properties render SharpEdge's simplicity, deployment-friendliness, and a wide scope of applicability. To the best of our knowledge, none of the prior work meets all these 4 properties (detailed in section~\ref{sec_related}). For instance, all IEEE 802.1Qav~\cite{cbs}, IEEE 802.1Qbv~\cite{tas}, IEEE 802.1Qch~\cite{cyclic}, and IEEE 802.1Qcr~\cite{ats} standards require changes in all switches in the network (not meeting property \#2). Moreover, IEEE 802.1Qav cannot support burstiness (not meeting property \#3) and cannot guarantee delay when link utilization is high (not meeting property \#1). Although IEEE 802.1Qbv~\cite{tas} and IEEE 802.1Qch~\cite{cyclic} can provide delay guarantee, they require fully synchronized network devices (not meeting property \#4). 

\textbf{Next sections}: The rest of this paper is organized as follows. First, in section~\ref{single_delay}, we consider a single bounded-delay for all traffic in a network and consider two cases for driving the theorems: 1) a single switch model of the network and 2) a multihop model of the network. Later, in section~\ref{multi_delay}, we consider the general case of having multiple bounded-delays for different classes of traffic in the network. Then, using the results of these two sections, we introduce SharpEdge in section~\ref{sec_sharp}. In section~\ref{sec_related}, we compare SharpEdge with various other schemes and after having a brief discussion in section~\ref{sec_dis}, we conclude the paper.

\section{Networks with single deterministic delay bound: The Theory}
\label{single_delay}
To start the analysis, we first distinguish two different sources of delay in the network. The overall delay that a packet experiences in the network consists of two parts:  1) the intrinsic/natural delay of the network ($T_n$) and 2) the variable delay caused by interfering with other packets in the network ($T_q$).

First part of the delay, $T_n$, includes serialization delay of a packet at each hop (i.e., transmission time of all bits in a packet over a finite link bandwidth), propagation delay on each link, and intrinsic processing times in the network devices. $T_n$ of a packet $p$ is equal to the delay that packet $p$ experiences in the network when the network is empty of other packets and only carries packet $p$. Clearly, $T_n$ depends on the architecture of the network, its topology, and its devices. In other words, for a given network, $T_n$ is a fixed number. 

Second part of the delay, $T_q$, is the only part of the delay that can be controlled for a given network. In other words, if we can show that $T_q$ can be bounded, then we have shown that the overall delay of a packet ($T_q+T_n$) in a network is bounded, because as mentioned earlier, for a given network $T_n$ is fixed. It can be seen that by definition, $T_q$ is the queuing delay of the network.

Therefore, in this paper, we mainly focus on queuing delay of the network, $T_q$, and the upper-bound of it. So, whenever we say delay experienced by a packet in a network, we mean $T_q$ unless it is specified otherwise.

\subsection{Part I: The Single Switch Model}
\label{sec_def}
Here, we first start with a single switch model of the network and derive the necessary and sufficient conditions to have a bounded-delay network. Later, we extend the model to the multi-hop scenario and introduce our main Theorem.

\begin{figure}[!t]
\centering
\includegraphics[width=0.7\linewidth,height=1.7in]{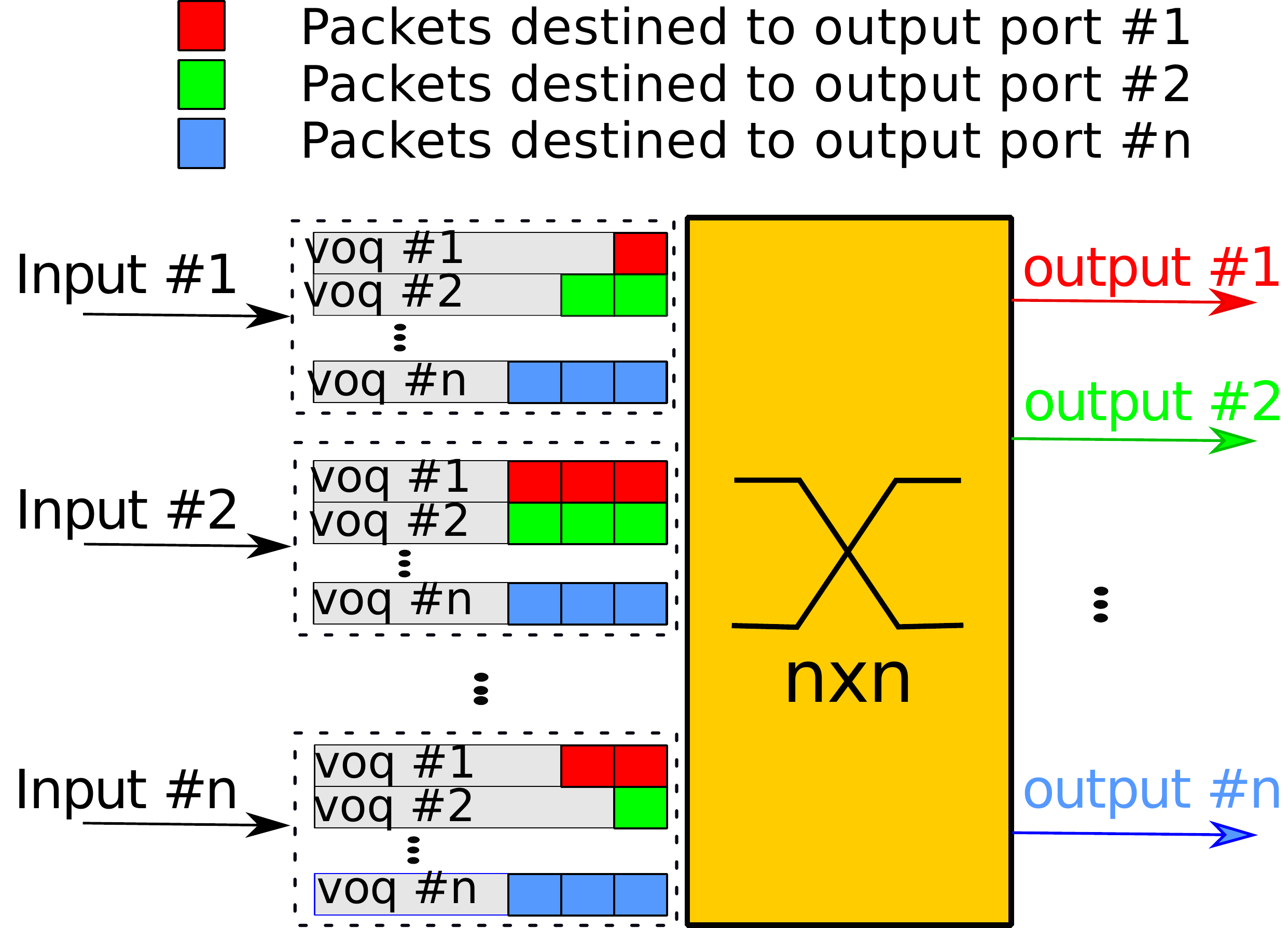}
\caption{Single VOQ switch model of the network}
\label{fig_model}
\end{figure}

In this section, we model the network with a single virtual output queue (VOQ) switch shown in Fig.~\ref{fig_model}. 

Without loss of generality, we assume that a packet is received when its last bit is received. Also, we assume that a packet is sent when its last bit is sent by the switch.

\subsubsection{Definitions}
\begin{define}
The sojourn-time of a packet ($T_q$) in a switch is the time from receiving it to the time that it is going to be sent (i.e., queuing delay not including service time).
\end{define}
\begin{define}
The transmission-time of a packet ($T_n$) is the time difference between when its first bit is being served/sent to when its last bit is served/sent (packet serialization time or service time).\footnote{Throughout this paper we assume that the propagation time and processing time in a device is negligible. In other words, $T_n$ dominates by the transmission time.} 
\end{define}
\begin{define}
The network has a bounded-delay of D when the maximum summation of sojourn-time and transmission time experienced by any incoming packet under any circumstances is bounded to D.
\end{define}
\begin{define}
For any arbitrary time t, BusyIndex(t) of a system 
is the total number of bits 
%coming to the the system during the time-window $[t, t+D)$.
awaiting to be served in the system at time t.
%\subsubsection{Definitions}    
\end{define}
\begin{define}
BI of a system is the maximum BusyIndex(t) of a system ($0\leq t<\infty$) .
%\subsubsection{Definitions}    
\label{def_bi}
\end{define}
\begin{define}
For any arbitrary time t and any arbitrary delay of d, InVolume(t,d) of the system is defined as the total number of bits coming or existed in the system during $[t, t+d)$.
%\subsubsection{Definitions}    
\label{def_bi}
\end{define}
\begin{define}
For any arbitrary time t and any arbitrary delay of d, OutVolume(t,d) of the system is defined as the total number of bits served by the system during $[t, t+d)$.
%\subsubsection{Definitions}    
\end{define}
\begin{define}
For any arbitrary time t and any arbitrary delay of d, Q(t,d) of the system is defined as the total number of bits queued in the system during $[t, t+d)$.
%\subsubsection{Definitions}    
\end{define}
\begin{define}
The maximum size of a packet in the network is $p^{max}$.
%\subsubsection{Definitions}    
\end{define}

%\textbf{BusyPeriod(t,D)}: For any arbitrary time t and any arbitrary delay of D, the maximum duration during the time-window $[t, t+D)$ that a switch serves packets uninterruptedly is called the BusyPeriod(t,D) of the network.
%
%Using the above definition, the bounded delay of a network can be expressed as BusyPeriod(0,$\infty$).
Now, using the single-switch model of the network and above definitions, we derive the necessary and sufficient conditions of having a bounded-delay network.

\subsubsection{Necessary Condition}
What is the minimal property for a bounded-delay network? We answer this question by the following theorem:
 
\begin{thm}
If the network has a bounded delay of D, for every arbitrary time t,
we must have:
\label{theo_nec}
\end{thm}
\begin{equation}
BI\leq D\times C
\label{eq_nec}
\end{equation}
where C is the capacity of the output link.

\begin{IEEEproof}
Proof by contradiction. Assume that the network has a bounded-delay of D while we have $BI>D\times C$. %Now, let's assume that the maximum number of bits waiting to be served in the system during $[t, t+D)$ happens to be at time $t'$ ($t\leq t' < t+D)$. 
Consider time $t$ when $BusyIndex(t)=BI$. At the time $t$, among the BI number of bits waiting to be served, there exists a bit which will be served last. Consider the packet $p$ corresponding to that bit. So, packet $p$ at least experiences $\frac{BI-S_p}{C}$ delay caused by other packets, where $S_p$ is the size of packet $p$. In addition, packet $p$ will experience a delay of $T_n=\frac{S_p}{C}$ for serving all its bits. So, for the sojourn-time of packet $p$ ,$T_q$, we will have: 
\small
\begin{equation}
\frac{BI-S_p}{C}\leq T_q
\Rightarrow        
\frac{BI}{C}\leq T_q + T_n
\Rightarrow        
BI\leq (T_q + T_n)\times C 
\end{equation}
\normalsize

On the other hand, the network has a bounded-delay of D for all packets including the packet $p$ i.e., $T_n+T_q \leq D$. Therefore, we have:
\small
\begin{equation}
BI\leq (T_q + T_n)\times C 
\Rightarrow
BI\leq D\times C
\end{equation}
\normalsize
which contradicts our assumption and proves Theorem~\ref{theo_nec}.  
\end{IEEEproof}

\begin{coro}
\label{cor_admission}
If the network has a bounded delay, then the input traffic must be admissible. 
\end{coro}
The intuition behind Theorem~\ref{theo_nec} is that in a bounded-delay network, the total traffic in the network is bounded at all times.

\subsubsection{Sufficient Condition}
\begin{figure}[!t]
\centering
\includegraphics[width=0.7\linewidth,height=1.5in]{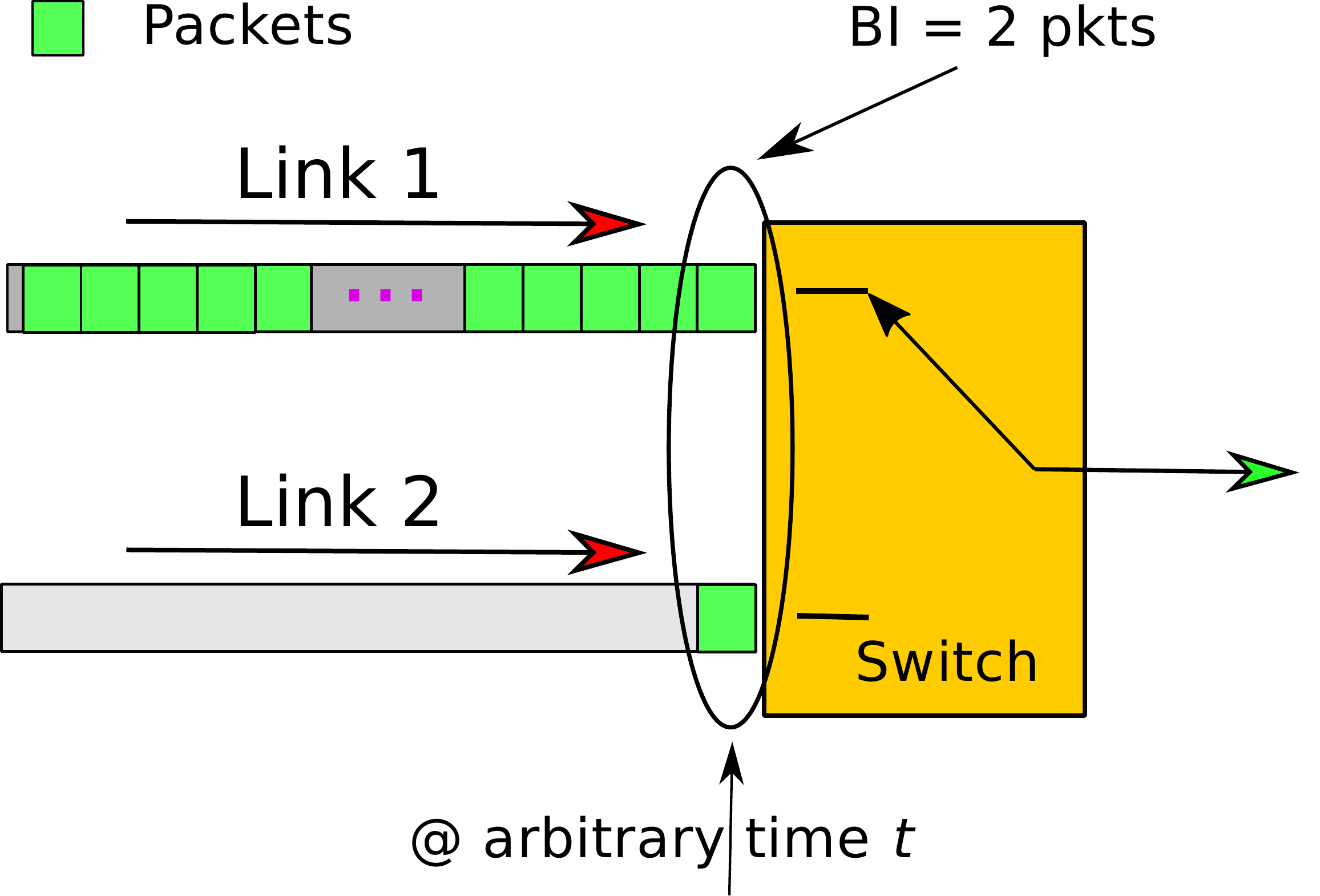}
\caption{A network with a bounded BI which does not have bounded-delay}
\label{fig_eg1}
\end{figure}

Notice that Equation~\ref{eq_nec} does not show a sufficient condition for having a bounded-delay network. In other words, a network can have a bounded BI, while not having a bounded-delay. For instance, consider the network shown in Fig.~\ref{fig_eg1}. Assume that the switch always serves the packets coming from the link 1, before serving packets from link 2. So, in the state shown in Fig.~\ref{fig_eg1}, BI of the system is always bounded to two packets while clearly packet on link 2 experiences sojourn-time of $\infty$.

So, to answer the question of ``Under what conditions will the network have a bounded delay of D?'', we introduce Theorem~\ref{theo_suf}.

\begin{thm}
If the network is work-conserving and if for any arbitrary time t, InVolume(t,D)$\leq$ D$\times C$, then the network will have the bounded delay of D.
%If for any arbitrary time t, $BusyIndex(t,D) \leq D\times C$ , then using any work-conserving  (scheduling/queue management) algorithm at the switch leads to having the bounded delay of D for the network.
\label{theo_suf}
\end{thm}
\begin{IEEEproof}
Proof by contradiction. Assume that for at least one packet, $p$, $T_q+T_n$ is larger than D, while the network is work-conserving and for any arbitrary time $t$, InVolume(t,D) $\leq$ D$\times C$.

Now, choose arbitrary time $t$ to be the time when the packet $p$ is received ($t_r$) and consider that $D<T_q+T_n$. Hence: 
\small
\begin{equation}
\text{InVolume}(t_r,D) \leq D\times C
%\Rightarrow
%\frac{\text{InVolume}(t_r,D)}{C}< \Delta
\label{eq_suf1}
\end{equation}
\normalsize
Since the network is work-conserving, and because during $[t_r,t_r+T_n+T_q]$ there exists at least one packet in the network (packet $p$), the network should have been busy serving packets uninterruptedly at least for the entire duration of $[t_r,t_r+T_q+T_n]$. Since overall delay of packet $p$ is larger than D, during $[t_r,t_r+T_q+T_n)$ at least $D\times C+1$ bits are in the network (1 bit represents the last bit of packet $p$ which by assumption, will be served after $t_r+T_q+T_n$). So, we have:
\small
\begin{equation}
\text{InVolume}(t_r,D) \geq D\times C+1
\end{equation}
\normalsize
which contradicts Equation ~\ref{eq_suf1} and proves Theorem~\ref{theo_suf}. 
\end{IEEEproof}

\begin{coro}
For a work-conserving network, using only proper shapers to bound the InVolume of the network is sufficient to reach a guaranteed deterministic delay bound for the network.
\label{cor_suf}
\end{coro}

\subsection{Part II: Multihop Model}
Corollary~\ref{cor_suf} indicates that using shapers to bound the InVolume at the edge of a single-switch model of the work-conserving network is sufficient for achieving the bounded-delay network. However, it is not clear whether this is necessarily the case in a multi-hop scenario. The key concern in a multi-hop scenario is the fact that the shaped traffic at the edge of a multi-hop network can lose its shape throughout the network. Therefore, it is a reasonable question to see whether Corollary~\ref{cor_suf} can be extended to the multi-hop case. To that end, we will introduce Theorem~\ref{theo_edge}.  

\begin{figure}[!t]
\centering
\includegraphics[width=0.7\linewidth,height=1.7in]{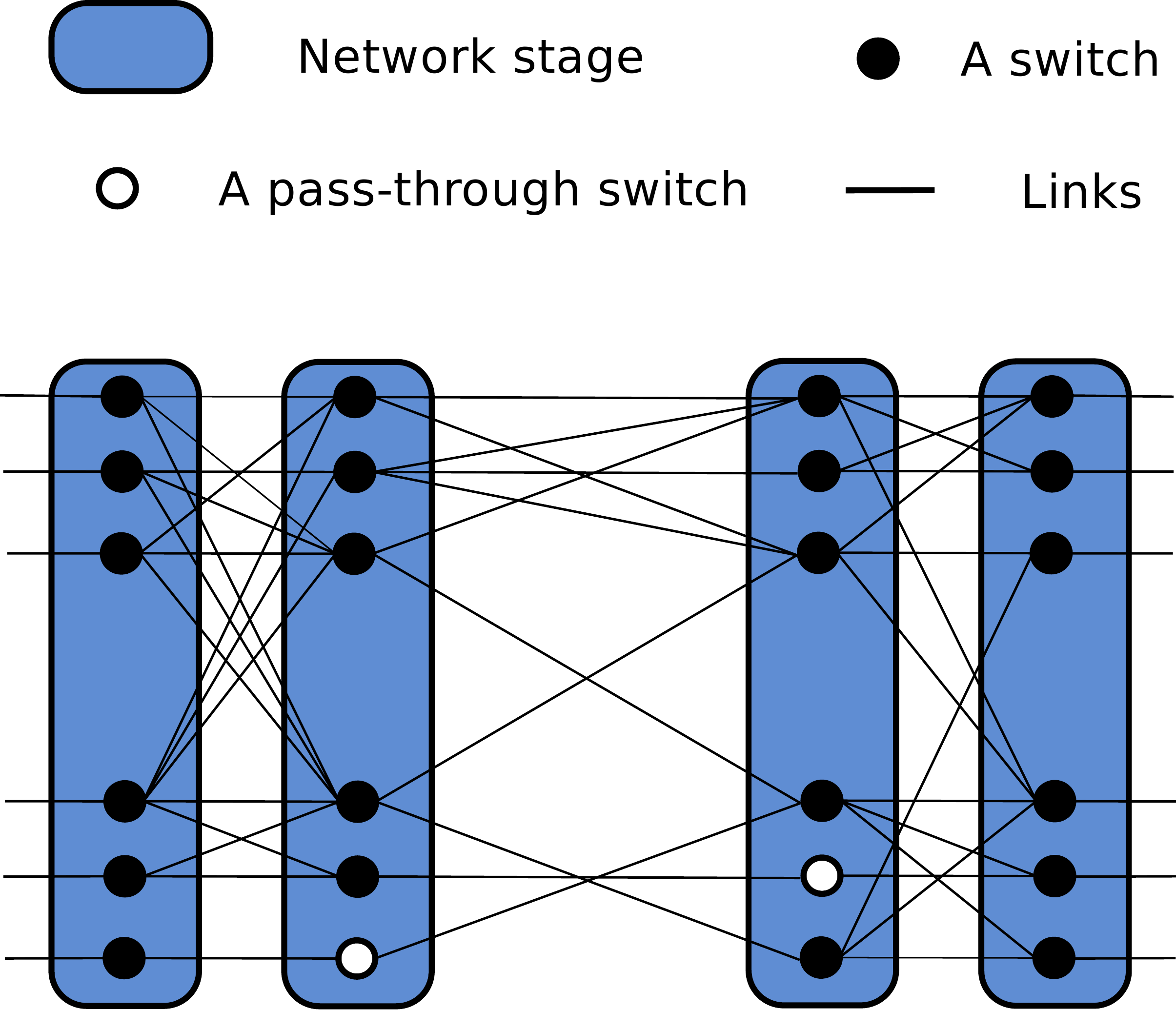}
\caption{Model of a multi hop network}
\label{fig_multi-model}
\end{figure}

Before that, we first model a network with maximum H hops as an H stage graph where input traffic passes through different stages and finally goes out of the network similar to the Fig.~\ref{fig_multi-model}. In a practical scenario, all flows are not necessarily passing through the same number of hops in the network. So, to make the model general and include all scenarios, we use imaginary pass-through switches in the model. In other words, if a source is connected to its destination through 4 switches in the network, while the maximum number of hops in the network is 5, we put an imaginary pass-through switch on one of the links in the path from this source to its destination. 
Hence, each node in the graph represents either a real switch or an imaginary switch and we can assume that all traffic will pass through the H stages. 

%To bound the overall delay of the network to D, we can bound the delay of each stage $i$ in the multi-hop model to $D_i$ subject to $\sum_{i=1}^{H} D_i=D$. Note that each stage can be modeled as a single-switch network. Therefore, Theorem~\ref{theo_suf} indicates that by bounding the InVolume(t,$D_i$) for stage $i$, we can achieve bounded-delay for the stage $i$ and therefore for the entire network. So, a clear sufficient condition for making bounded-delay network is to put shapers at the ingress of each stage. But can we do better than that?

Notation: To use the single-switch definitions mentioned in section~\ref{sec_def} here, we use superscript $(h)$ to specify the stage number. For example, $\text{InVolume}(t,D)^{(h)}$ becomes the total number of bits coming or existed in the stage $h$ during $[t,t+D)$.

%\begin{lem}
%For work-conserving stages $h$ and $h+1$ in the network, we have :
%if for any arbitrary time $t$ $\text{InVolume}(t,D)^{(h)}\leq D\times C$, then: % for any arbitrary time $t$: 
%\footnote{Here, for simplicity, we assumed that the propagation and processing time at stages are zero. When propagation and processing time combined is $\epsilon$, there will only be a constant time shift which does not impact the lemma (i.e., $\text{InVolume}^{(h)}(t,D) = \text{OutVolume}^{(h)}(t+\epsilon,D)$).}
%\begin{equation}
%%\text{InVolume}^{(h)}(t,D) = \text{OutVolume}^{(h)}(t,D)
%\text{OutVolume}^{(h)}(t,D) = \text{InVolume}^{(h+1)}(t,D)
%\label{eq_suf_edge}
%\end{equation}
%\end{lem}

%Proof by contradiction. Assume that for any arbitrary time $t$ we have $\text{InVolume}^{(h)}(t,D)\leq D\times C$ but we don't have $\text{InVolume}^{(h)}(t,D) = \text{OutVolume}^{(h)}(t,D)$. So, there should be $\text{InVolume}^{(h)}(t,D)>\text{OutVolume}^{(h)}(t,D)$ for at least one $t$. \footnote{Notice that the system only serves bits and it does not generate them, therefore we cannot have $\text{InVolume}^{(h)}(t,D)<\text{OutVolume}^{(h)}(t,D)$.} 

\begin{thm}
To achieve a bounded-delay network, it is sufficient to shape the traffic properly only at the edge of the network and use any work-conserving algorithm at the switches to send packets throughout the network. 
\label{theo_edge}
\end{thm}

\begin{IEEEproof}
There are two parts of the delay for any incoming packet: $T_n$ and $T_q$. Assume that maximum number of hops in the network is H (it is clear that $H<\infty$). Therefore, for any packet, we have $T_n\le H\times \frac{p^{max}}{C}$, where C is the capacity of the bottleneck link. Now, to obtain a bound on $T_q$, we exclude $T_n$ from the calculations. By definition, at an arbitrary stage $h$ in the network we have: \footnote{Note that by definition, $\text{InVolume}^{(h)}(t,D)$ includes both incoming and queued packets during $[t,t+D)$.}
\begin{equation}
Q^{(h)}(t,D)=\text{InVolume}^{(h)}(t,D)-\text{OutVolume}^{(h)}(t,D)
\label{eq_q}
\end{equation}
On the other hand, it is clear that the output of each stage is the input of the next stage, so: \footnote{Here, for simplicity, we assumed that the propagation and processing time at stages are zero. When propagation and processing time combined is $\epsilon$, there will only be a constant time shift in the equation~\ref{eq_out-in} (i.e., $\text{OutVolume}^{(h)}(t,D) = \text{InVolume}^{(h+1)}(t+\epsilon,D)$).}
\begin{equation}
\text{OutVolume}^{(h)}(t,D)=\text{InVolume}^{(h+1)}(t,D)
\label{eq_out-in}
\end{equation}
Now, we calculate the total bits queued in the system during $[t,t+D)$, Q(t,D). By definition:
\begin{equation}
Q(t,D)=\sum_{h=1}^{H} Q^{(h)}(t,D)
\end{equation}
Using Equations~\ref{eq_q} and ~\ref{eq_out-in} we have:
\begin{equation*}
Q(t,D)=\sum_{h=1}^{H} (\text{InVolume}^{(h)}(t,D)-\text{OutVolume}^{(h)}(t,D)) 
\end{equation*}
\begin{equation}
= \text{InVolume}^{(1)}(t,D)-\text{OutVolume}^{(H)}(t,D)
\label{eq_q-total}
\end{equation}
Now, assume that the traffic is shaped \textit{properly} at the edge of the network to satisfy the following equation for \textbf{any} arbitrary time $t$ and a given delay of D:
\begin{equation}
\text{InVolume}^{(1)}(t,D)\leq D\times C
\label{eq_edge_suf}
\end{equation}
where C is the capacity of the bottleneck link. Also, since network itself does not generate packets, we have:
\begin{equation}
\text{InVolume}^{(h)}(t,D)\geq\text{OutVolume}^{(h)}(t,D) 
\end{equation}
Therefore:
\begin{equation}
0\leq Q(t,D) \leq D\times C
\label{eq_q-total}
\end{equation}

An arriving packet will be served at most after Q(t,D) bits have been served in the network. In addition, network stages are assumed to be work-conserving. Hence, the maximum delay that any bit experiences due to the all queued bits in the network, will be:
\begin{equation}
% \text{Max}(T_q)=
\frac{\text{Max}(Q(t,D))}{C}=D
\label{eq_q-dmax}
\end{equation}
Now consider the bit which experiences the maximum queuing delay in the network, and its corresponding packet, $p$. The overall delay of packet $p$, both overall transmission and queuing delays, will be (Note that the transmission time of any packet at the first stage is already calculated in $\frac{\text{InVolume}^{(1)}(t,D)}{C}$): 
% So, for any packet, we have:
% \small
\begin{align}
\label{eq_q_n_multi}
    \sum_{i=1}^{H}(T_q^{(i)}+T_n^{(i)})=&(T_n^{(1)}+\sum_{i=1}^{H}(T_q^{(i)})+\sum_{i=2}^{H}T_n^{(i)}\le  \nonumber\\
    &\le (D)+((H-1)\times\frac{p^{max}}{C})
\end{align}
\normalsize
% \begin{equation}
% \le D+(H-1)\times\frac{MTS}{C}
% \end{equation}
So, the Theorem is proved. 
\end{IEEEproof}

\textbf{Notice 1}: Note that a naive delay bound for the network can be derived by simply adding a worst-case delay (D) at each stage and achieve $H\times D$ as the delay bound of the network. We call this delay bound the greedy delay bound of the network. In contrast with the greedy delay bound (which usually is used as the base for calculation of worst-case delay in the literature (e.g. as in~\cite{sg1,sg2,cbs,tas,cyclic,ats})), Theorem~\ref{theo_edge} provides a worst-case delay for the network independent of the number of hops. 

\textbf{Notice 2}: Also, note that Theorem~\ref{theo_edge} guarantees a bounded-delay for the network independent of the switching/scheduling algorithms used throughout the network. 

\section{Networks with Multiple Deterministic Delay Bounds: The Theory}
\label{multi_delay}
Up to this point, we assumed that there is only one class of traffic and one deterministic delay-bound for the network. However, in the real world, there will be multiple services/classes each requiring different delay-bounds. Therefore, we extend our analysis to provide multiple delay bounds for multiple classes in this section. 

\begin{figure}[!t]
\centering
\includegraphics[width=0.7\linewidth,height=1.9in]{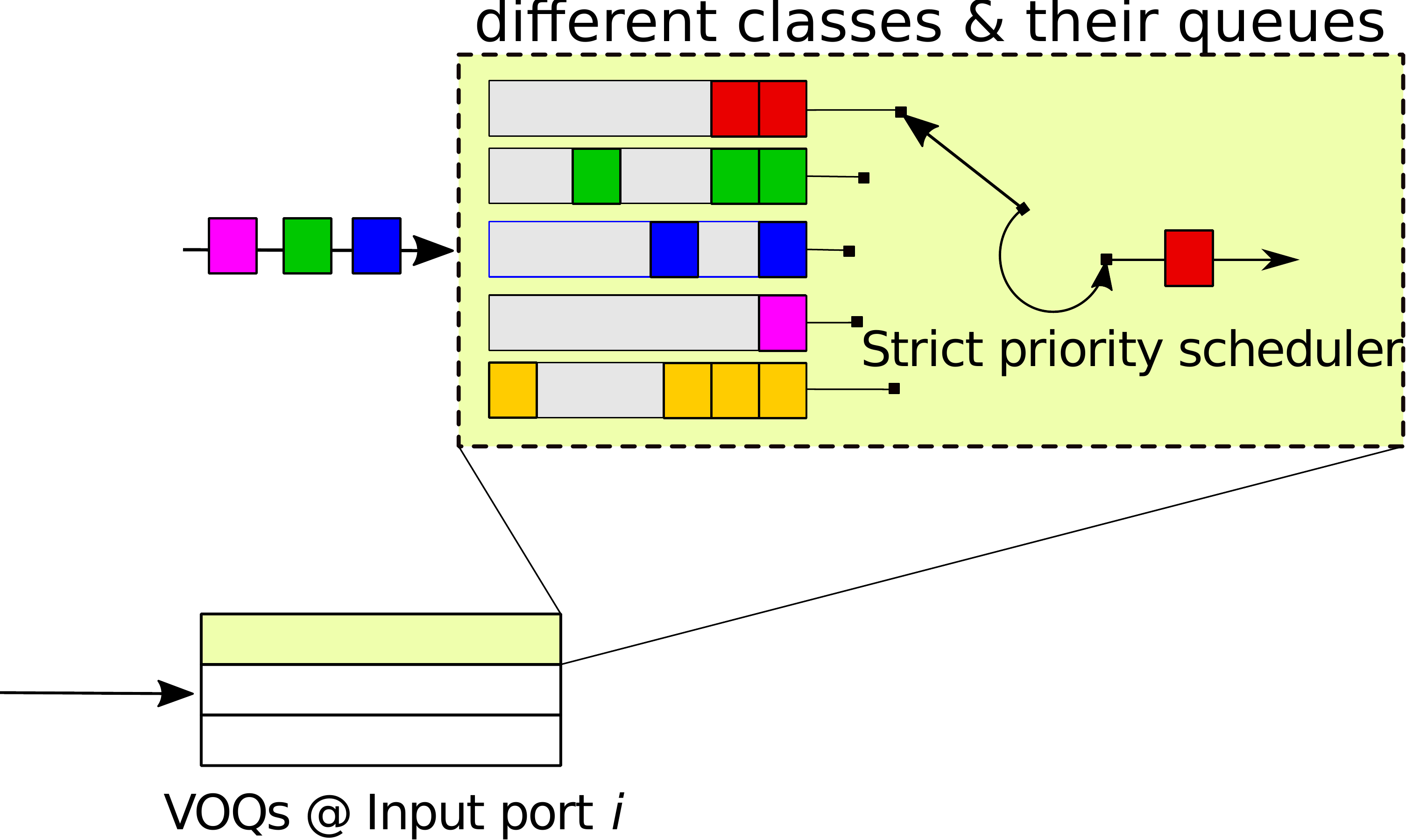}
\caption{multi-class}
\label{fig_multi-class}
\end{figure}

One of our key objectives is to not change the core switches in the network. So, we use the simple non-preemptive strict priority mechanism available in commodity switches (and used in Diffserv proposal~\cite{diffserv}) to deal with the multiple service requirements. The multi-priority queue structure is shown in Fig.~\ref{fig_multi-class}. %to support the multiple delay bounds for the network.  
At the ingress, packets are classified into different queues based on their tagged classes/priorities. Later, queues will be served in a non-preemptive strict priority manner. In other words, a packet from a lower priority queue can be served only when all higher priority queues are empty. In addition, the lower priority packet that is being served by the system cannot be preempted by a recently received higher priority packet. Now, using this architecture, we update our previous analysis.

\subsection{Definition}
Here, to use the single delay bound definitions mentioned in section~\ref{sec_def}, we use subscript $(k)$ to specify the class/priority number. For example, $\text{InVolume}(t,D)^{(h)}_{(k)}$ becomes the total number of bits of class $k$ coming or existed in the stage $h$ during $[t,t+D)$. The notations used in this section are summarized in Table~\ref{table_sum}.

\begin{table}[!h] \renewcommand{\arraystretch}{1} \caption{Notations} 
\label{table_sum} \centering
\begin{tabulary}{\columnwidth}{c|C}
\hline
Notation & Description \\
\hline
$H$	& maximum number of hops \\
\hline
$K$ & The total number of classes with bounded-delay requirements\\
\hline
$P_{(k)}^{max}$ & The maximum packet size of flows in class $k$ \\
\hline
$n_{(k)}$ & The number of flows in class $k$ \\
\hline
$C_{(k)}$ & The bandwidth associated to class $k$ on bottleneck link\\
\hline
$D_{(k)}$ & The desired bounded delay of the class $k$\\
\hline
%$\sigma_{i(k)}$ & The total bits that flow $i$ in class $k$ allowed to send at any arbitrary time-window of size $D_{(k)}$ \\
\hline
\end{tabulary}
\end{table}
We say the network $N$ has $K+1$ non-preemptive strict-priority queues, when it satisfies all the followings:
\begin{enumerate}
\item There are totally $K+1$ classes including $K$ classes of traffic with different delay-bound requirements (class 1 to $K$) and one class of best-effort traffic (class $K+1$).
\item Each switch in the network consists of $K+1$ queues with different priorities where highest priority ($k=1$) is given to the smallest delay-bound traffic and lowest priority ($k=K+1$) is given to the best-effort traffic.
\item A packet from a certain class can be served only if no packet belonging to any higher priority class exists (strict-priority policy). 
\item A packet that is being served cannot be preempted by other packets (non-preemptive policy).
\end{enumerate}

\subsection{Sufficient Condition}
Here, we update Theorem~\ref{theo_edge} to reflect multiple priority classes with different delay-bounds. 

\begin{lem}
If for any arbitrary time $t$:
\begin{equation}
\text{InVolume}(t,d)^{(h)}_{(k)}\leq B
\label{eq_lem-bound}
\end{equation}
Then, for any $d'=a\times d$ ($1<a\in\mathbb{Z}$), we have: %\in\Re$):
\begin{align}
\text{InVolume}(t,d')^{(h)}_{(k)}=a\times \text{InVolume}(t,d)^{(h)}_{(k)}\leq a\times B
\label{eq_lem-res}
\end{align}
\label{eq_lem}
\end{lem}

\begin{IEEEproof}
When $a\in\mathbb{Z}$, time period $[t,t+d')$ can be divided into $a$ sub-periods of size $d$. So, we have $\text{InVolume}(t,d')^{(1)}_{(k)}=a\times \text{InVolume}(t,d)^{(1)}_{(k)}$. %Also, when $a\notin\mathbb{Z}$, we have $\text{InVolume}(t,d')^{(1)}_{(k)}\leq (a+1)\times \text{InVolume}(t,d)^{(1)}_{(k)}$. 
Now, using the upper bound shown in Equation~\ref{eq_lem-bound} for $\text{InVolume}(t,d)^{(1)}_{(k)}$ leads to Equation~\ref{eq_lem-res}. 
\end{IEEEproof}

\begin{thm}
If network $N$ has $K+1$ non-preemptive strict-priority queues and it satisfies the following conditions:
\begin{enumerate}
\item The network is work-conserving.
\item For any arbitrary time t and $1\leq k\leq K$:
\begin{equation*}
\text{InVolume}(t,d_{(k)})^{(1)}_{(k)}\leq d_{(k)}\times C_{(k)}  
\end{equation*}
\item For $K$ classes with delay-bound requirements:
\begin{equation*}
%\small
\sum_{k=1}^{K}C_{(k)}\leq C
\end{equation*}
\item For $i\leq k\leq K$: 
\begin{equation*}
\left(\frac{d_{(k)}}{d_{(i)}}\right)\in\mathbb{Z}
\end{equation*}
\end{enumerate}

Then, packets in class $k$ ($1\leq k\leq K$) will have the total bounded delay of $D_{(k)}=\Delta_{(k)}+\epsilon_{(k)}+\zeta_{(k)}$, where:

\begin{align} 
%&\Delta_{(k)}=d_{(k)}+\sum_{i=1}^{k-1}\left[\left(\ceil*{\frac{d_{(k)}}{d_{(i)}}}-\floor*{\frac{d_{(k)}}{d_{(i)}}}\right)\frac{C_{(i)}}{C}\right]
&\Delta_{(k)}=\alpha_{(k)}d_{(k)} \text{\ \ \ \ \ \ } & (\alpha_{(k)}\leq 1)\label{eq_delta}\\
&\epsilon_{(k)}=\sum_{i=1}^{H}\max_{k<i\leq K+1}(\frac{P_{(i)}^{max}}{C})\label{eq_epsilon}\\
&\zeta_{(k)}=(H-1)\frac{P_{(k)}^{max}}{C}\label{zeta_epsilon}
\end{align} 
%
%D_{main}\leq d_{(k)}\times C_{(k)}  
%    \sum_{k=1}^{K}C_{(k)}\leq C
%\right. 
\label{theo_suf2}
\end{thm}

\begin{IEEEproof}
Consider a packet $p$, belonging to the class $k$ ($1\leq k\leq K$). The maximum delay experienced by packet $p$ will consist of three parts: 1) Delay caused by serving packets with higher or equal priorities in the system before packet $p$ is served ($\Delta_{(k)}$) , 2) Delay caused by serving a lower priority packet which started to receive service just before packet $p$ comes to the system ($\epsilon_{(k)}$)\footnote{due to the non-preemptive assumption about the system.}, and 3) Delay caused for transmission/serialization of packets at each stage. Now, we calculate these parts. 

Consider the network $N_k$ as the network made by removing all lower priority classes than $k$ from the network $N$. Note that by definition, the value of $\Delta_{(k)}$ for network $N$ is equal to $\Delta_{(k)}$ of the network $N_k$, because $\Delta_{(k)}$ for the packet $p$ only includes delays caused by the packets in classes with the same or higher priorities than $k$. 

In the network $N_k$, $\Delta_{(k)}$ is equal to the bounded-delay of network $N_k$ and based on Theorem~\ref{theo_edge}, the maximum value of $\text{InVolume}^{(1)}(t,d_{(k)})$ for network $N_k$ shows the bounded-delay of the network $N_k$. Now, we consider condition 4, Lemma~\ref{eq_lem}, and inequality of condition 2 to calculate the $\text{InVolume}^{(1)}(t,d_{(k)})$ of the network $N_k$ as follows:
\begin{align}
&\text{InVolume}^{(1)}(t,d_{(k)})=\sum_{i=1}^{k}\text{InVolume}^{(1)}_{(i)}(t,d_{(k)})\leq\nonumber\\
&\leq \sum_{i=1}^{k}\left({\frac{d_{(k)}}{d_{(i)}}}\times\text{InVolume}^{(1)}_{(i)}(t,d_{(i)})\right)\nonumber\\
&\leq \sum_{i=1}^{k}\left({\frac{d_{(k)}}{d_{(i)}}}\times d_{(i)}\times C_{(i)}\right)=d_{(k)}\sum_{i=1}^{k}C_{(i)}\nonumber
\end{align}
%&\leq \sum_{i=1}^{k}\left(\left(\ceil*{\frac{d_{(k)}}{d_{(i)}}}-\floor*{\frac{d_{(k)}}{d_{(i)}}}
Therefore, similar to the discussions leading to Equation~\ref{eq_q-dmax} in proof of Theorem~\ref{theo_edge}, we get: $\Delta_{(k)}={d_{(k)}\sum_{i=1}^{k}C_{(i)}}/{C}$.

Now, considering condition 3, we have:
\begin{align*}
\alpha_{(k)}=\frac{\sum_{i=1}^{k}C_{(i)}}{C}\leq 1 \Rightarrow &\Delta_{(k)}=\alpha_{(k)}d_{(k)} \text{\ } & (\alpha_{(k)}\leq 1)\\
\end{align*}

Which proves Equation~\ref{eq_delta}. To calculate $\epsilon_{(k)}$, note that in a non-preemptive regime, on each hop throughout the path of a packet $p$ of class $k$ in the network, at most one packet from lower priority classes than $k$ can get service just before the packet $p$ is received on that hop. So, for the packet $p$, the maximum interfering delay caused by lower priority classes will be calculated as in Equation~\ref{eq_epsilon}. 

On the other hand, similar to the discussions leading to Equation~\ref{eq_q_n_multi} in proof of Theorem~\ref{theo_edge}, the transmission delays of a packet after the first stage will be calculated by equation~\ref{zeta_epsilon}.  
\end{IEEEproof}
A more general form of Theorem~\ref{theo_suf2} can be derived by removing the restriction enforced by condition 4 (i.e., assuming $\frac{d_{(k)}}{d_{(i)}}\in\mathbb{R}$) and replacing $\Delta_{(k)}$ in Equation~\ref{eq_delta} with the following: 
\begin{align} 
%&\Delta_{(k)}=\sum_{i=1}^{k}d_{(k)} \text{\ \ \ \ \ \ } & (\alpha_{(k)}\leq 1)\label{eq_delta_gen}\\
%&\epsilon_{(k)}=\sum_{i=1}^{H}\max_{k<i\leq K+1}(\frac{P_{(i)}^{max}}{C})\\
\Delta_{(k)}=\alpha_{(k)} d_{(k)}+\sum_{i=1}^{k-1}\left(\ceil*{\frac{d_{(k)}}{d_{(i)}}}-\floor*{\frac{d_{(k)}}{d_{(i)}}}\right)\frac{C_{(i)}}{C}
\end{align} 

Due to space limitation, we leave the proof of the general form of Theorem~\ref{theo_suf2} to our future work. 
%This will not impact discussions of the our next 
%and the fact that the more general form of Theorem~\ref{theo_suf2} does not impact 
%\vspace{2in}

\section{The SharpEdge}
\label{sec_sharp}
Any design that satisfies Equation~\ref{eq_edge_suf} in Theorem~\ref{theo_edge} (and in general, Theorem~\ref{theo_suf2}) will guarantee the bounded-delay for the network. In this section, we focus on presenting one of such a designs called SharpEdge. SharpEdge is an architecture directly motivated by the Equation~\ref{eq_edge_suf}. In SharpEdge, edge of the network shapes the traffic according to the Theorem~\ref{theo_edge} so that a bounded-delay for the network can be guaranteed. SharEdge uses a novel shaping scheme called Quantum shaper. In what follows we describe the Quantum shaper algorithm.

%\begin{figure}[!h]
%\centering
%\includegraphics[width=0.9\linewidth,height=1.3in]{figs/shaper}
%\caption{The Big picture of SharpEdge System}
%\label{fig_big}
%\end{figure}
\subsection{The Quantum Shaper}
\begin{figure}[!t]
\centering
    \begin{minipage}[b]{\linewidth}
	\centering
     \includegraphics[width=0.5\linewidth,height=1in,keepaspectratio]{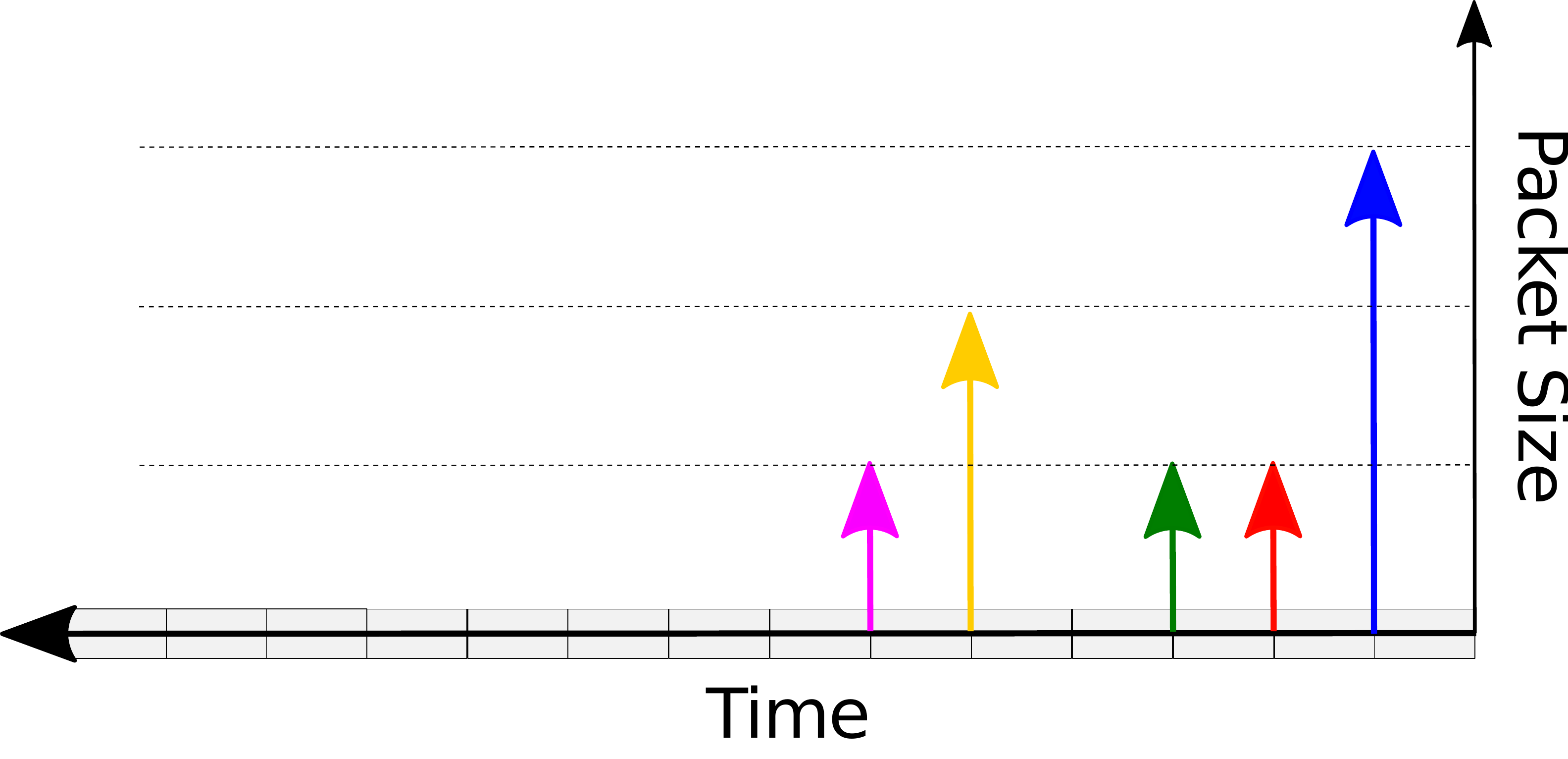}
	\subcaption{Packets at ingress}	
	\label{fig_shaper_ingress}
	\end{minipage}
	\hfill
    \begin{minipage}[b]{\linewidth}
	\centering
     \includegraphics[width=0.8\linewidth,height=1.5in]{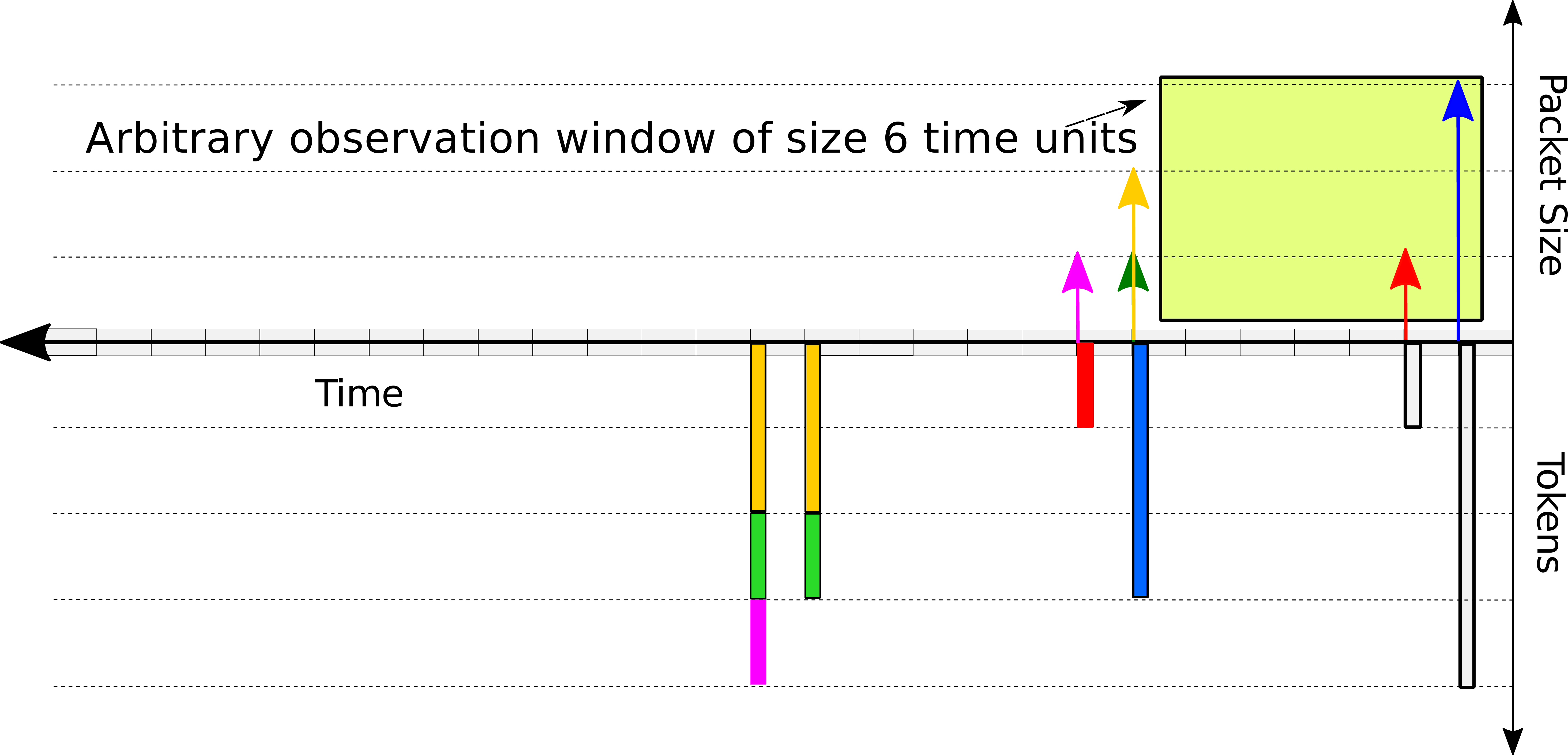}
	\subcaption{Packets at egress of a Quantum shaper (the bounded units of data during arbitrary time window of 6 units is guaranteed)}
	\label{fig_shaper_quantum}
	\end{minipage}
	\hfill
    \begin{minipage}[b]{\linewidth}
	\centering

     \includegraphics[width=0.8\linewidth,height=1.5in]{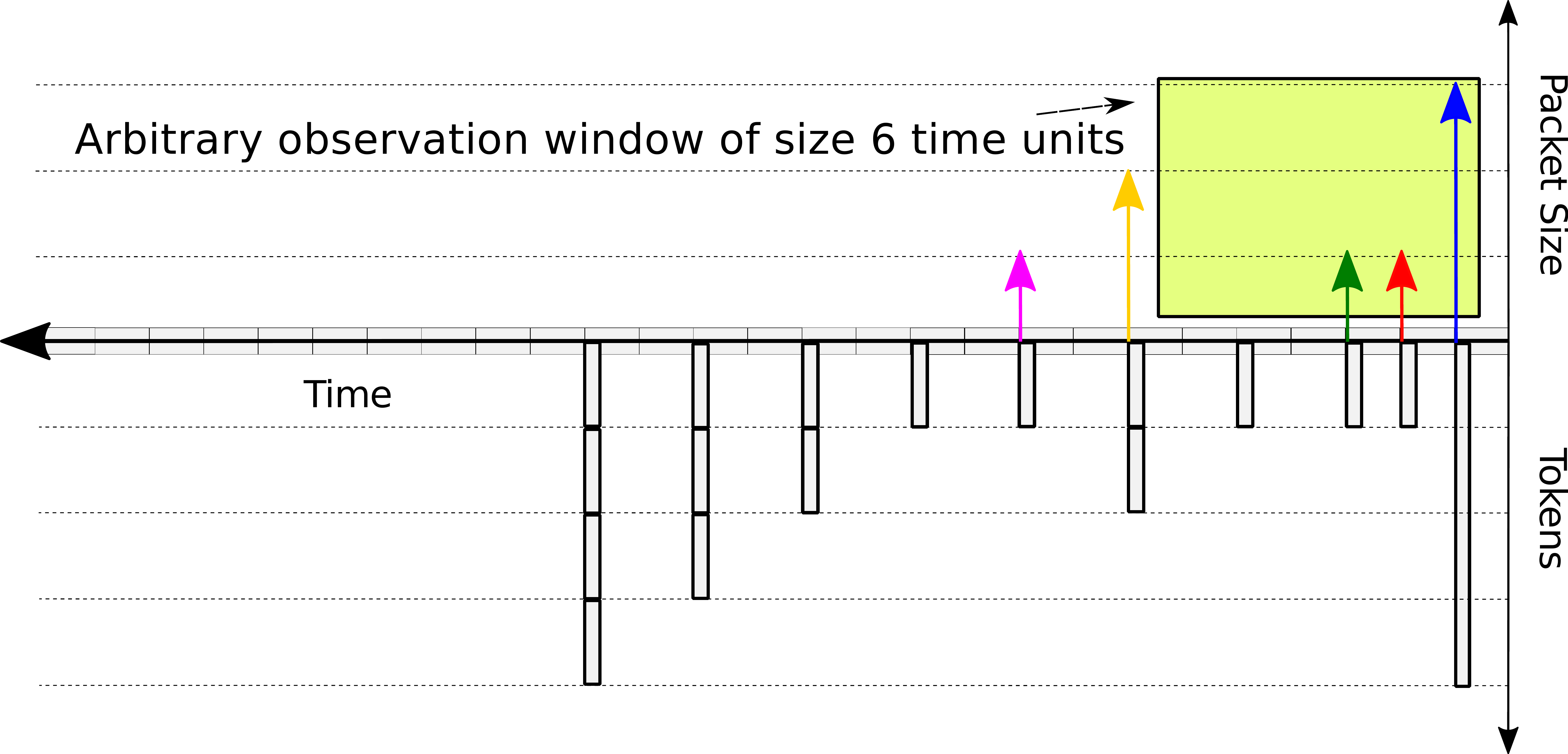}
	\subcaption{Packets at egress of a leaky bucket with repl. rate of 3 unit of pkt per 6 time unit (the restriction of having bounded units of data during shown time window is violated)}
	\label{fig_shaper_lky3}
	\end{minipage}
	\hfill
    \begin{minipage}[b]{\linewidth}
	\centering
     \includegraphics[width=0.8\linewidth,height=1.5in]{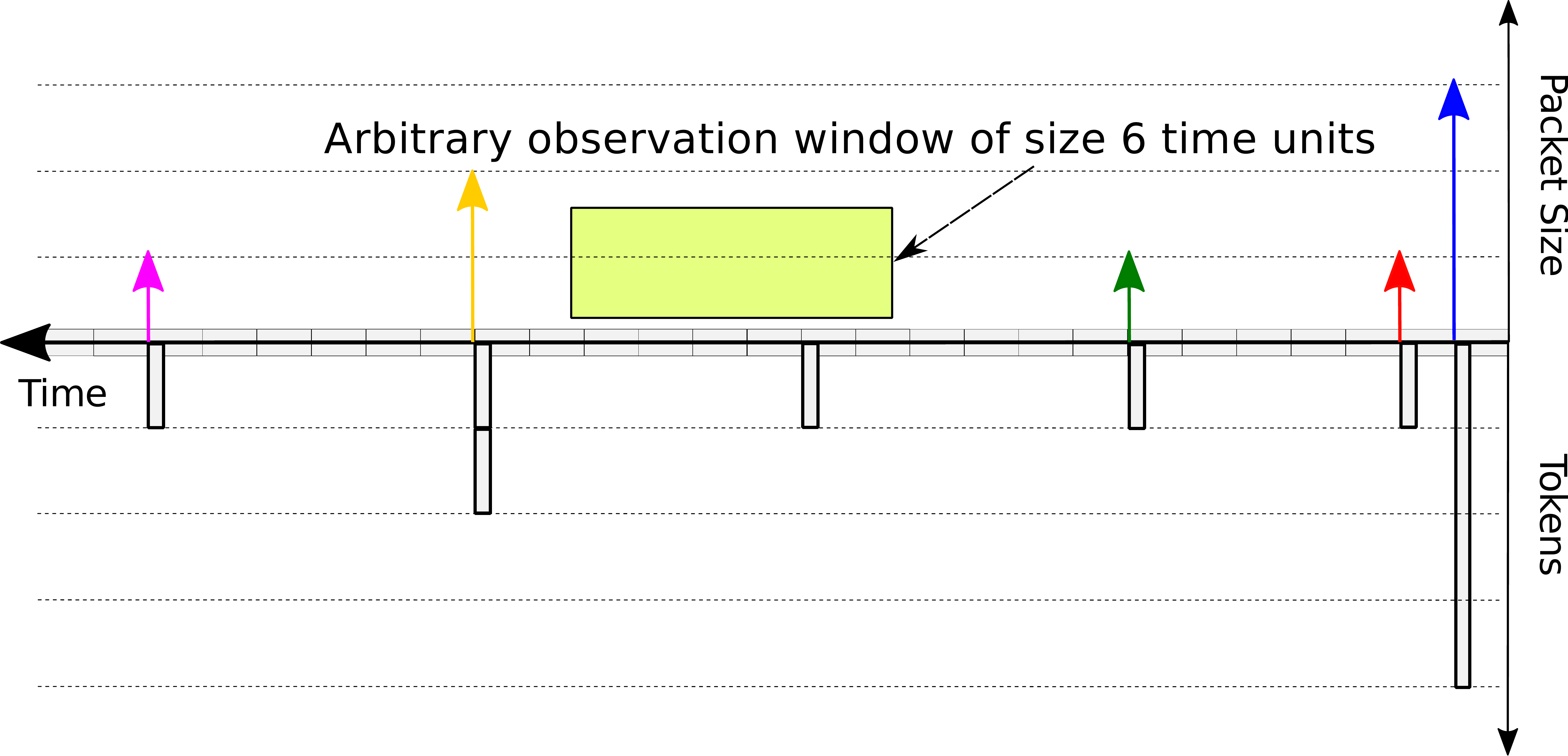}
	\subcaption{Packets at egress of a leaky bucket with repl. rate of 1 unit of pkt per 6 time unit}
	\label{fig_shaper_lky1}
	\end{minipage}
	\caption{Big picture of Quantum Shaper and comparison with Leaky bucket. In this scenario, One flow sends packets with different sizes to the network. The arrows in the figures show individual packets of the flow where the height of each arrow represents the size of packets. On the other hand, rectangles represent tokens.}
\label{fig_shaper}
\end{figure}  
\begin{figure}[!h]
\small
 \removelatexerror
  \begin{algorithm}[H]
  \DontPrintSemicolon
   \caption{The Quantum Shaper's Pseudo Code}
   \label{alg_shaper}
	
	\Fn(){Shaper()}{ 
		\tcc*[l]{Initialization}
		$Credit(0)=\sigma$\;
		\While{true}{
			$t_{now} \longleftarrow$ Current time\;
			\tcc*[l]{Check if new packets can be sent}
 			SendNextPacket()\;
		}
	}
	\setcounter{AlgoLine}{0}
	\Fn(){SendNextPacket()}{ 
        \tcc*[l]{Check if you can send a packet}
%        $avg\_rtt \longleftarrow$ \textit{average rtt during previous Tuning Cycle}\;
%        $min\_\alpha \longleftarrow 1$\;
%        $max\_\alpha \longleftarrow 10$\;
		\uIf(){$Q.size>0$}{ \tcp*[l]{Peak HOL packet}
           	$p=Q.head(); S_p= p.length;$\;
            
%            $\alpha\mathrel{+}= \frac{Target-avg\_rtt}{2avg\_rtt}$\;
            	\uIf(){$Credit(t_{now})>S_p$}{ %\tcc*[l]{}
            		$Credit(t_{now})\mathrel{-}=S_p$\;
	                \textbf{Output} $p$\;
	                \tcc*[l]{Schedule an increase in the credits at the time $t_{now}+D$}
     		           IncreaseCredit$(t_{now}+D, S_p)$\;
            	}
		}
	}
	\setcounter{AlgoLine}{0}
	\Fn(){IncreaseCredit(t,credit)}{ 
%      	\textbf{@wait until time t, then:} \;
      		$Credit(t)\mathrel{+}=credit$\;
      		$Credit(t)=max(Credit(t),\sigma)$\;
			\tcc*[l]{When no credit is consumed or repl. during $(t_1,t_2)$, for any $t'$ $(t_1<t'<t_2)$ we have: $Credit(t')=Credit(t_2)=Credit(t_1)$}
      }
	
%	        \uElseIf{$Target < avg\_rtt$}{%\tcc*[l]{waiting phase}
%            $\alpha\mathrel{-}= \frac{2(avg\_rtt-Target)}{Target}$\;
%            \uIf(){$\alpha \leq min\_\alpha$}{ %\tcc*[l]{}
%                $\alpha=min\_\alpha$\;
%            }
%        }
%     }
  \end{algorithm}
\end{figure}
%Now, we propose a novel shaper algorithm called quantum shaper to satisfy Equation~\ref{eq_edge_suf}. 
Assume that for a given network (with maximum number of hops of $H$) the desired bounded-delay of the network is $D'$ and $D=D'-\frac{(H-1)p^{max}}{C}$. If for any arbitrary time $t$, the total bits that the flow $i$ can send to the network during $[t,t+D)$ is limited to $\sigma_i$ and if Equation~\ref{eq_sigma} holds, then Theorem~\ref{theo_edge} indicates that the network will have the bounded-delay of $D'$ (Note that $\text{InVolume}^{(1)}(t,D)\leq \sum_{i=1}^{n}\sigma_i$ where n is the number of incoming flows).
\begin{equation}
\sum_{i=1}^{n}\sigma_i \leq D\times C
\label{eq_sigma}
\end{equation}

Therefore, the key question becomes ``how can we shape the flow's traffic to bound its total bytes during \textbf{any} arbitrary time-window of size D?''. We answer this question by introducing Quantum shaper.

Consider a bucket of credits with a maximum size of $\sigma_i$ credits. An incoming packet of size $P$ requires consuming $P$ credits from the bucket to be permitted for being released to the network. The bucket is full of $\sigma_i$ credits at the beginning. When $P$ credits are used by an incoming packet at time $t$, all $P$ consumed credits will be replenished at time $t+D$ in a ``quantized'' manner. In other words, in contrast with a leaky bucket shaper~\cite{leaky} where credits are replenished with a fixed rate in a continues manner through time, in Quantum shaper, credits are reproduced only at certain times in the future. This quantized nature enables the Quantum shaper to make the total bits of a flow $i$ entering the network always bounded to $\sigma_i$ for any arbitrary time-window of size D. The Quantum shaper's pseudocode is shown in Algorithm~\ref{alg_shaper}.

To show that Quantum shaper can bound the total bits in the network for any arbitrary time-window of size D, we use the following example. Consider that a flow sends packets with variable sizes at different times as shown in Fig.~\ref{fig_shaper_ingress}. Assume that the desired $D$ (delay-bound associated to the class which the flow belongs to) is 6 time units and $\sigma_i=4$ unit for this flow. At time 0, there are $\sigma_i=4$ (units) credits in the bucket. A (blue) packet of size 3 units coming at time 1 consumes 3 unit of credits. All these 3 units of credits will be replenished at time $1+6$. At time 2, a (red) packet of size 1 unit consumes 1 remaining credit. So, at time $2+6$ a new credit will be inserted in the bucket. From time 3 to time 6, there is no credit in the bucket, so no packet will be sent out. At time 7, packets with sizes of 2 and 1 units will be sent out and later at time 8, the (pink) packet with size 1 unit will be sent. This is shown in Fig.~\ref{fig_shaper_quantum}. As this example illustrates, using Quantum shaper, the total number of packets of flow $i$ is always guaranteed to be bounded to $\sigma_i$ in any arbitrary time window of size D and this as described earlier, leads to the overall bounded delay of D for the system.

To show the difference between credit replenishment in the Quantum shaper and the leaky bucket, we use two leaky buckets with replenishment rates of 3 and 1 packet units per 6 time units to shape the example shown in Fig.~\ref{fig_shaper_ingress}. The output packets are illustrated in Fig.~\ref{fig_shaper_lky3} and Fig.~\ref{fig_shaper_lky1}, respectively. The continuous replenishment of the credits in leaky bucket leads to violation of the bounded number of bytes during an arbitrary time window of size 6 (as shown in Fig.~\ref{fig_shaper_lky3}). This problem can be resolved by reducing the replenishment rate (as in Fig.~\ref{fig_shaper_lky1}), but the reduction of replenishment rate leads to underutilization issue.

\section{Related Work and Comparison with SharpEdge}
\newcommand{\blue}{\cellcolor{LighterBlue}} 
\newcommand{\vio}{\cellcolor{Vio}} 
\newcommand{\org}{\cellcolor{Org}} 
\newcommand{\g}{\cellcolor{LightGreen}} 
\newcommand{\x}{\textcolor{Red}\xmark} 
\newcommand{\gc}{\textcolor{Green}\cmark} 
\newcommand{\cs}{\textcolor{Green}{\cmark*}} 
\begin{table*}[!t] \renewcommand{\arraystretch}{1.2} \caption{Comparison among different schemes (Considering the core network), (Check the text for more information on the fields with *)} 
\center
\label{t_com} \centering
\begin{tabular}{|l | c | c | c| c| c| c| c| c| c| c}
\hline
                                            &                     \multicolumn{4}{c|}{\org Performance} & \multicolumn{3}{c|}{\g Deployment Friendliness}\\
    Scheme                                &    Delay         &    Support     & Work                & No greedy	& No Per-Flow  & No switch & No time\\
                                            &    Guarantee &    Burst         &    Conserving         & delay-bound& management&modification & synchronization\\
%                                            &     &             &             &  &   &             & \\
\hline                                                                                                                                                    
    \blue SharpEdge                        & \blue \gc         &    \blue \gc    &    \blue \gc            &\blue \gc     &\blue \gc                & \blue \gc                & \blue \gc           \\
\hline                                                                                                                                                  
    IEEE 802.1Qav~\cite{cbs}            & \x*             &    \x            &    \x                    & \x        &    \gc                    &    \x                    &    \gc                 \\
    IEEE 802.1Qbv~\cite{tas}            & \gc            &    \gc            &    \x                    & \gc        &    \gc                    &     \x                    &    \x                    \\
    IEEE 802.1Qch~\cite{cyclic}            & \gc            &    \gc            &    \x                    & \x        &    \gc                    &    \x                    &    \x                    \\
    IEEE 802.1Qcr~\cite{ats}            & \gc             &    \gc            &    \x                    & \x        &    \x                    &     \x                    &    \gc                     \\    
\hline                                                                                                                                                  
    Intserv~\cite{intserv}                & \gc             &    \gc            &    \x                     & \gc        &    \x                     &     \x                     &    \gc            \\
    Diffserv~\cite{diffserv}                & \x              &    \gc            &    \gc                    & -         &    \gc                 &     \gc                 &     \gc                \\
\hline                                                                                                                                                  
    GPS/PGPS~\cite{gps1,gps_multi}    & \gc             &    \gc            &    \gc                    & \gc        &    \x                     &   \x                     &    \gc                 \\
    Stop-and-Go~\cite{sg1,sg2}            & \gc             &    \gc            &    \x                    & \x        &    \gc                    &    \x                 &    \gc                  \\
    HRR    ~\cite{hrr}                        & \gc             &    \gc            &    \x                    &     \x        &    \gc                    &    \x                    &    \gc        \\
    D-EDD    ~\cite{edd_d}                & \gc             &    \gc            &    \gc                    & \x        &    \gc                    &    \x                     &    \gc             \\
%    Virtual Clock~\cite{vc}                & ?                 &    ?            &    ?                    &             &    ?                     &    \x                    &    ?            \\
    RCSP~\cite{rcsp}                        & \gc             &    \gc            &    \x                    &      \x       &    \x                     &    \x                    &    \gc          \\
%     AFDX~\cite{afdx}                        & \gc            &    \x            &    \gc                    &             &    \gc                 &     \x                   &    \gc            \\
    BoDe~\cite{bode}                    & \gc            &    \gc            &    \gc                    & \x         &    \gc                 &    \x                    &    \gc               \\
\hline                                                                                                                                                 
    D3~\cite{d3}                            & \gc             &    \gc            &    \gc                    & \gc        &    \x                     &    \x                    &    \gc          \\
    pFabric~\cite{pfabric}                & \x             &    \gc            &    \gc                    & -            &    \x                     &    \x                    &    \gc            \\
    PDQ~\cite{pdq}                        & \gc             &    \gc            &    \gc                    & \gc        &    \x                     &    \x                    &    \gc            \\
    Qjump~\cite{qjump}                    & \x*            &    \gc            &    \gc                    & -         &    \gc                    &    \gc                 &    \gc                  \\
    HyLine~\cite{hyline}                    & \x            &    \gc            &    \gc                    & -         &    \gc                 &    \gc                    &    \gc               \\
%    TDMA Ethernet~\cite{eth_tdma}    & \gc            &    \x            &    \gc                    &             &    \x                     &     \cs                  &                        \\
\hline                                                                                                                                                 
    Vegas~\cite{vegas}                            & \x             &    \gc            &    \gc                     & -            &    \gc                   &    \gc                  &    \gc        \\
    Sprout~\cite{sprout}                            & \x             &    \gc            &    \gc                     & -            &    \gc                   &    \gc                  &    \gc        \\
    NATCP~\cite{natcp}                              & \x             &    \gc            &    \gc                    & -            &    \gc                   &    \gc                  &    \gc       \\
    C2TCP~\cite{c2tcp,c2tcp2}                        & \x             &    \gc            &    \gc                    & -            &    \gc                   &    \gc                  &    \gc           \\
    DeepCC~\cite{deepcc}                & \x             &    \gc            &    \gc                    & -            &    \gc                   &    \gc                  &    \gc             \\
    Orca~\cite{orca}                & \x             &    \gc            &    \gc                    & -            &    \gc                   &    \gc                  &    \gc             \\
\hline
\hline
\end{tabular}
\end{table*}

\label{sec_related}
Here, we briefly mention some of the related work and compare SharpEdge with them. The summary of the comparisons made are gathered in Table~\ref{t_com}. \\
The bottom line is that SarpEdge is the only scheme which can guarantee multiple delay bounds for multiple classes of traffic, support burstiness, requires no time synchronization, and does not need any changes in the core of the network.
\subsection{General Proposals}
\subsubsection{IntServ~\cite{intserv}} Integrated service (IntServ) is an extension proposal for Internet architecture proposed in early 90s. It defines a service called guaranteed service which targets a perfectly reliable upper bound on delay. IntServ introduces the primary requirement of per-flow management and packet scheduling (Fair-Queuing) in the network. However, the implementation complexity of Fair-Queuing made Intserv an impractical approach toward providing delay guarantees in the network. 

\subsubsection{DiffServ~\cite{diffserv}} Impractical nature of IntServ led to a simple proposal by Cisco called differentiated service (DiffServ). The key idea in DiffServ was to avoid the per-flow management of IntServ in the network by putting all flows into a limited number of classes and treat all flows in a class in the same way in the network. However, the simplicity of DiffServ came with the cost of no delay guarantee for the packets.

\subsubsection{IEEE 802.1CM~\cite{8021cm}}
IEEE 802.1CM or Time-Sensitive Networking for Fronthaul standard~\cite{8021cm} is a joint effort by CPRI and IEEE 802.1 groups to provide bridged Ethernet connectivity for fronthaul networks. The big picture of IEEE 802.1CM standard is that it tries to describe fronthaul requirements, specify two classes of traffic, and mention synchronization requirements for the bridges to ease support of time sensitive flows. However, it suffers from two main points. First, as it is declared in the Clause 8.1, the calculation of the end-to-end delay provided in IEEE 802.1CM is a "greedy bound" calculation meaning it simply uses the very conservative approach of multiplying the worst-case delay at each hop by the number of hops. Second, it specifies token bucket as the requirement for shaping/policing the traffic. In contrast, as we showed in this paper, queuing delay of the system is independent of the number of the hops and leaky/token bucket shaping is inefficient compared to the Quantum shaper (e.g. see discussion around Fig.~\ref{fig_shaper}).

\subsection{Queue Management and Scheduling Proposals}
\subsubsection{GPS~\cite{gps1,gps_multi}} GPS and its packetized version (PGPS) are among the classic packet scheduling designs which use the concepts of Fair-Queuing and traffic shaping to bring rate/delay performance guarantees. GPS shows that when input flows are shaped using leaky buckets~\cite{leaky}, weighted fair-queuing can bound the maximum delay of packets in the network.  Although GPS is an ideal fair-queuing scheme, per-flow management and fair-queuing implementation of GPS/PGPS remains the key practical issue of it. 
\subsubsection{Stop-and-Go~\cite{sg1,sg2}} Stop-and-Go was one of the first approaches that used the time-framing concept to send traffic throughout the network. In Stop-and-Go, the time axis is divided into separate frames with size $T$ seconds and only data that is received in the previous frame is allowed to be sent out in the current frame. Stop-and-Go shows that if all input flows are shaped in a proper way, it can bound the delay of packets to $2\times T\times H$, where H is the number of hopes. In other words, it simply uses a greedy bound on the delay. In contrast, SharpEdge provides delay bound independent of the number of hops (a non-greedy delay bound). In addition, Stop-and-Go is a non-work conserving approach which can lead to underutilized links. 
\subsubsection{RCSP~\cite{rcsp}} Rate Controlled Strict Priority (RCSP) is a non-work-conserving queuing framework which decouples the rate control block from the scheduling block to provide performance guarantees. Per-flow rate regulators in RCSP employ the idea of virtual departure time~\cite{vc} and assign an eligibility time for the departure of an incoming packet. At the calculated time, it releases the packet to the next stage, i.e., the scheduler and the scheduler considers the priority assigned to each flow for serving them.

\subsection{Queue/Resource Management Proposals for Datacenters} 
\subsubsection{D3~\cite{d3}} 
D3 was one of the earliest designs in datacenters targeting deadline-aware communications in the network. D3 intends to meet flow deadlines. It does not have a notion of per packet delay bound. More importantly, results of using D3 highly depends on the flow arrival, because it allocates bandwidth to flows in a first-come-first-serve manner. That means a flow with higher priority can simply miss its deadline if it comes after a flow with lower priority to the network.Also, D3 uses a special switch and NIC hardware and modifies end-hosts to have a new transport protocol. 

\subsubsection{pFabric~\cite{pfabric}} 
pFabric replaces FIFO queues with priority queues where packets from different flows are sorted using a priority tag set by the sender. pFabric can reduce the average completion time of flows but it cannot guarantee a delay-bound for the network. Also, pFabric suffers from the complexity and high cost of implementing priority queues in the switches.  

\subsubsection{Qjump~\cite{qjump}} 
Qjump employs DiffServ to classify flows and it attempts to rate-limit the traffic of each class at end-hosts to reduce the interference delay of a class on other classes. However, Qjump can only bound the delay of the highest priority class in the network. Also, to prevent the need for time synchronization of devices, Qjump restricts the utilization of the highest priority class to half of the ideal utilization that it can achieve. Particularly, when bottleneck link bandwidth is $C$ and desired bounded delay is $D$, there can be at most $C/D$ flows each sending 1 packet every $D$ seconds to the network. However, Qjump restricts each flow to send only one packet every $2\times D$ seconds. Doing that, it can prevent the need for time synchronization between the devices, though this dramatically degrades the overall achieved utilization. For instance, when the start time of the timeframes at end hosts is the same, the total utilization gained by Qjump will be at most 50\%. That being said, even when only the highest priority class is considered (the class that Qjump can provide it with a bounded-delay), SharpEdge can achieve two times higher total utilization compared to Qjump.

\subsection{Transport Layer Solutions} 
Some recent transport control protocol (TCP) designs target satisfying ultra-low latency packet transmissions in the network (e.g. \cite{sprout,natcp,c2tcp,c2tcp2,deepcc,orca}). 
%Four of the most recent delay-oriented TCP designs are Copa~\cite{copa}, PCC-Vivace~\cite{vivace}, C2TCP~\cite{c2tcp}, and ExLL~\cite{exll}. 
In the big picture, the common property of all these TCP designs is that they attempt to reduce the delay of packets in the network. As fully end-to-end approaches, most of the TCP designs do not require any change in the network. However, TCP designs cannot provide any guarantee for the delay of packets. Due to TCP's acknowledgment based structure, TCP protocols are basically not work-conserving; however, when packets are sent to the network, they will be served in a work-conserving manner in the core network. That is because these TCPs do not require network device modifications and current general commercial network switches are work-conserving. 

\subsection{Network Calculus} 
Through years different tools have been used to model a network and analyse it. 
% For instance, one of the popular approaches is to model the network with a fluid-model as done in GPS\cite{gps1,gps_multi}. 
One of these tools is network calculus (NC) which has been used for analyzing performance guarantees in computer networks for years. However, nearly all of the work in this context follows strict and usually the same assumptions to achieve the results. For instance, works that use NC to achieve delay-bounds for the network usually assume that the input traffic follows a leaky-bucketed form. Then, using this simplified model, they can show under certain strict assumptions about the schedulers or AQM schemes used in the network (e.g. a simple tail-drop FIFO or a more complicated earliest deadline first scheduler), they can achieve a "greedy delay-bound" for the network, which is of the form $H\times D$, where $H$ is the number of hops and $D$ is the maximum delay on a hop~\cite{nc_book1,nc_book2,nc_simple1,nc_simple2}. 

Compared to the works using NC-based models, our analysis has three very important different properties. First, our analysis is based on no assumption about the schedulers or other algorithms that might be employed in the switches and we have shown that the delay-bound is actually independent of them. Second, the delay-bound that we presented here, is not a "greedy delay-bound". That means in the delay-bound that we present, the entire summation of the queuing delay that a packet can experience in a network is independent of the number of hops in that network. Third, as we showed in Fig.~\ref{fig_shaper}, although a leaky-bucketed traffic is simple to analyze and greatly simplifies the NC/fluid-model analysis, use of the leaky-bucket in practice leads to under utilization issues.

\subsection{Others}
Also, there is another set of works that are dedicated to delay analysis of the existing proposals. In other words, these works assume that a \textit{specific} scheme is used in the network and try to come up with a closed-form of delay-bound for the network. Among these works, some recent examples such as~\cite{ubs,ubs_analysis2,ubs_analysis3} attempt to analyze the dealy-bound of a network when UBS scheme~\cite{ubs} (which is currently used as the base for IEEE 802.1Qcr amendment) is used in all switches in the network. For example, authors in \cite{ubs} study a delay-bound for a single-class one-hop network. Authors in \cite{ubs_analysis2} extend the analysis and propose a max-plus representation of UBS. Later, authors in~\cite{ubs_analysis3} use these analyses and study the impact of CBS on the delay. 

Although these analyses can be helpful to show that a \textit{specific} scheme can bound the delay, they all miss an important point. The point is that they basically forget the key problems of the specific underlying scheme that they are analyzing. For instance, as we discussed in section~\ref{sec_802.1Qcr}, 802.1Qcr (and basically UBS), requires per-flow shaping (per-flow management) at each hop in the network which means all switches in the network are required to be changed. Also, 802.1Qcr simply follows the greedy delay bound approach in which the end-to-end delay is calculated by considering a maximum delay per hop and then multiplying that by the number of hops which means, similar to IEEE 802.1Qch, the overall delay will depend on the number of hops. Moreover, 802.1Qcr is a non-work-conserving design that leads to underutilized links.

In contrast, in this paper, we perform our analysis without tying it to any specific scheme. By identifying key \textit{general} \textit{simple} and \textit{intuitive} principles and theorems for having a bounded-delay network, we pave the way to introduce more practical and simpler designs. Our analysis, in contrast with all mentioned works, achieves a straight-forward and intuitive delay-bound while being transparent to the underlying switches. Using these general theorems, we introduced a sample design named SharpEdge which can guarantee a bounded delay for the network while not having the issues of other existing solutions.  

% All these works go through complicated delay analysis of a \textit{specific} scheme which usually cannot go beyond being a mathematical formulation of the 
% do not  any and end-up with some closed-form of delay-bound for the network  they missed an important point. The point is that whether the underlying scheme that they are trying to analyse is good 

% as we mentioned in section~\ref{sec_802.1Qcr}, IEEE is currently working on an asynchronous solution under the name of 802.1Qcr amendment and uses UBS~\cite{ubs} scheme as the base for 802.1Qcr. Although as we discussed in section~\ref{sec_802.1Qcr}, 802.1Qcr have requires per-flow shaping (per-flow management) at each hop in the network which means all switches in the network are required to be changed. Second, 802.1Qcr simply follows the greedy delay bound approach in which the end-to-end delay is calculated by considering a maximum delay per hop and then multiplying that by the number of hops. So, similar to IEEE 802.1Qch, the overall delay will depend on number of hops. Moreover, RCSP (therefore, 802.1Qcr) is a non-work-conserving design which leads to underutilized links.

\section{Discussion}
\label{sec_dis}
\subsection{Where the shapers will be implemented?}
The precise answer to this question depends on the nature of the network itself. For instance, when end-hosts are managed by the network provider (e.g. in datacenter networks, industrial networks, local networks in an organization, etc.), shaping of the traffic can be enforced by the servers~\footnote{For instance, in case of having virtual devices, the Hypervisor is a very good candidate to enforce the shaping}. Another different example is the cellular network where the end-hosts are not necessarily managed by the network provider. In these cases, edge switches/devices are proper candidates for enforcing the shaping.

\subsection{Combination of Shaping and Policing at the Edge}
When shaping is done at the end-hosts, edge of network will police the traffic to make sure that a malicious user does not impact the delay of other users by sending more than what it is allowed to send. Therefore, combination of shaping and policing at the edge will protect unwanted interference from malicious users and help keeping delay bounded.

\subsection{Are the current network switches work-conserving?}
Scheduling packets from input links to output links of VOQ-based switches might lead to occasional internal blocking due to the imperfect scheduling algorithms used in the switches. Internal blocking of a switch means that packets at the inputs are not scheduled to be sent to their corresponding idle outputs. This makes the system non-work-conserving, though the idle time of the system (the time period when output ports are idle, though there are packets on input ports) would be very small. 
Today's switches use different techniques to resolve the possible internal blocking and imperfect nature of scheduling algorithms including speedup, expanded buffering, etc. (\cite{highperfromance, speedup1,speedup2,speedup3,speedup4,speedup5,arista710,arista728,aristacloud,ciscomeraki}).  
%For instance, Broadcom switches use speedup of 2 to prevent internal blocking issue~\cite{?}. 
After using these techniques, there might still be very small idle time at the output. However, this delay would be negligible considering today's multiple Gbps link speeds and target bounded delays of 10s of ms. Therefore, it can be safely considered that today's switches are work-conserving devices.

\subsection{Practical Deployment Issues}
As stated in Corollary~\ref{cor_admission}, admission control is a have-to requirement for achieving bounded-delay networks. That means in contrast with a best-effort medium that users  do not follow any constraints and send traffic as much as they want to the network, all input traffic to a bounded-delay network is required to pass an admission control phase. In this paper, we assumed that admission control has already been done and restrictions on each class of traffic have already been calculated. In practice, this task might not be straight forward and practically performing admission control in SharpEdge (or any other architecture) is an empirical concern that we leave to our future work. The fact is that providing delay-guarantee is a new service and supporting this new service might add to the complexity of the network and its protocols. However, here, an important note is that since SharpEdge does not require any changes in the core of network, its admission control solution will potentially be significantly simpler than admission control in an architecture where all switches are required to be involved in the process.  

The big picture is that a bounded-delay guarantee for each packet by a network provider is a special service and users who are interested in this service are required to identify their needs through communications with the service providers either in an online/on-demand or offline manner. Through this communication, network provider can update its admission control and shaping policy rules corresponding to the contract with that user at the edge devices. Then, when users are registered to certain classes of service and the edge devices are updated, users can start sending their traffic to the network.

\subsection{What Next?}
In this paper we showed that satisfying the objective of bounding the delay of a network is independent of the choice of the scheduling schemes, AQM designs, etc. that may be used in the network. In other words, it is sufficient to shape the traffic \textbf{only} at the edge of the network to achieve a hard bounded-delay for the entire network. However, this may bring up the questions such as what is the difference between using various scheduling, AQM, or other algorithms in switches then? Here, a very important point is that we only considered the objective of having hard bound on the delay of packets and did not discuss the metrics such as delay-jitter, average delay of a flow, 90th percentile delay, etc. The algorithms employed in the switches such as packet scheduling and AQM differ from each other on achieving these extra objectives. For instance, controlling or bounding the 90th percentile delay of packets most likely requires the involvement of proper packet scheduling algorithms at each hop in the network and solely shaping traffic at the edge might not be the solution. Putting all together, although SharpEdge architecture guarantees the bounded-delay property of 
the network, satisfying more objectives, such as bounding 90th percentile packet delay, require more novel designs and improvements of the architecture.

\section{Conclusion}
In this paper, we revisited the problem of providing deterministic bounded-delay guarantees for the network. To that end, we investigated the theory behind hard bounded-delay networks and derived the necessary and sufficient conditions for achieving deterministic bounded-delays in the network. Based on the derived theorems, we proved that as long as a network is work-conserving, independent of the packet scheduling and queue management algorithms in network switches, it is sufficient to shape the traffic~\textit{properly} at the edge of the network to reach hard bounded-delays in the network. Using this, we presented SharpEdge, a simple deployment-friendly scheme which does not require any changes in the core of the network to guarantee multiple bounded-delays for a network. SharpEdge enables delay-sensitive traffic to meet their hard deterministic delay bounds while they can coexist with throughput-hungry traffic in the network. Also, SharpEdge is an asynchronous approach which does not require any time synchronization among network devices.

%\section*{Acknowledgments}
%Author(s) would like to thank anonymous individual \#1, \#2, and \#3 for their valuable discussions and feedback on the work presented here.  
% \vspace{1in}
% \pagebreak

%\bibliographystyle{IEEEtran}
%\bibliographystyle{spmpsci}
%\bibliography{ref}

\end{document}